\documentclass[reprint,
superscriptaddress,
amsmath,
amssymb,
aps]{revtex4-2}
\usepackage{multirow}
\usepackage{siunitx} 
\usepackage{braket}
\usepackage{mathtools}
\usepackage{graphicx}
\usepackage{color}
\usepackage[bookmarks=true,colorlinks=true,urlcolor=black,linkcolor=black,citecolor=black,breaklinks]{hyperref}

\usepackage{xspace}
\usepackage{amsfonts}
\usepackage{bm}

\newcommand{\moire}[0]{moir\'{e}\xspace}

\definecolor{pu}{RGB}{200,50,200}
\definecolor{gr}{RGB}{0,187,0}
\definecolor{bl}{RGB}{68,34,200}
\definecolor{re}{RGB}{200,34,68}
\definecolor{ye}{RGB}{255,165,0}
\definecolor{oran}{RGB}{255,170,0}

\begin{document}

\title{Theoretical determination of the effect of a screening gate on plasmon-induced superconductivity in twisted bilayer graphene}

\author{Liangtao Peng}
\affiliation{Department of Physics, Faculty of Science, National University of Singapore, 2 Science Drive 3, Singapore 117542}
\affiliation{Centre for Advanced 2D Materials, National University of Singapore, 6 Science Drive 2, Singapore 117546}
\author{Indra Yudhistira}
\affiliation{Centre for Advanced 2D Materials, National University of Singapore, 6 Science Drive 2, Singapore 117546}
\affiliation{Department of Physics, Faculty of Science, National University of Singapore, 2 Science Drive 3, Singapore 117542}
\author{Giovanni Vignale}
\affiliation{The Institute for Functional Intelligent Materials, National University of Singapore, 4 Science Drive 2, Singapore 117544}
\author{Shaffique Adam}
\affiliation{Centre for Advanced 2D Materials, National University of Singapore, 6 Science Drive 2, Singapore 117546}
\affiliation{Department of Physics, Faculty of Science, National University of Singapore, 2 Science Drive 3, Singapore 117542}
\affiliation{Department of Materials Science and Engineering, 
National University of Singapore, 9 Engineering Drive 1, 
Singapore 117575}
\affiliation{Yale-NUS College, 16 College Ave West, Singapore 138527}

\begin{abstract}
The microscopic pairing mechanism for superconductivity in magic-angle twisted bilayer graphene remains an open question. Recent experimental studies seem to rule out a purely electronic mechanism due to the insensitivity of the critical superconducting temperature to either a highly doped screening layer or the proximity to a metallic screening gate. In this theoretical work, we explore the role of external screening layers on the superconducting properties of twisted bilayer graphene within a purely electronic mechanism. Consistent with the experimental observations, we find that the critical temperature is unaffected by screening unless the screening layer is closer than 3 nanometers from the superconductor.  Thus, the available transport data is not in contradiction with a plasmon-mediated mechanism. We also investigate other properties of this plasmon-mediated superconductivity including signatures in the tunneling density of states as probed in spectroscopy experiments.
\end{abstract}

\maketitle

\section{Introduction}\label{sec:intro}

The discovery of superconductivity in magic-angle twisted bilayer graphene (MATBG)~\cite{Cao_Nature2018_correlated, Cao_Nature2018_unconventional} has attracted tremendous interest, in part due to the similarity of the observed phase diagram with the long-standing problem of high-temperature superconductivity in cuprates. At present, there is no consensus on the microscopic mechanism for superconductivity in a system as simple as two rotated sheets of carbon. A number of theoretical studies based on the Bardeen-Cooper-Schrieffer (BCS) approach have focused on a phonon-mediated mechanism~\cite{FengchengWu_PRL2018_Phonon, Bernevig_PRL2019_Phonon, Peltonen_PRB2018_Meanfield}. Typically, a pure phonon mechanism gives a critical temperature $T_c\sim 1\mathrm{K}$ (slightly below what is seen in experiments).  However, considering the dynamical polarizability, we have argued recently~\cite{Indra_PRB2019_Gaugephonon, Girish_NC2021_transport} that the phonon deformation potential is likely to be strongly screened by the large density of states in MATBG. In contrast, as was pointed out recently by~\cite{Guinea_PNAS2021_Coulomb, Lewandowski_PRB2021_umklapp}, Umklapp processes in the reduced moire Brillouin zone might act to increase the strength of phonon pairing. On the other hand, the emergence of flat bands~\cite{MacDonald_PNAS2011_Moireband} strongly enhances the electron-electron interaction which favors plasmonic superconductivity~\cite{Girish_PRR2020_Plasmon, Guinea_PNAS2021_Coulomb, Lewandowski_PRB2021_umklapp} with larger critical temperatures $T_c\sim10\mathrm{K}$. 

Recently, three experimental papers~\cite{Levitov_Natrue2020_Untying, Saito_NatruePhys2020_TBG, Xiaoxue_Science2021_BLG} have investigated the role of screening on this superconducting state. Refs.~\cite{Levitov_Natrue2020_Untying, Saito_NatruePhys2020_TBG} varied the distance of an external metal screening gate from 6.7 nm to 68 nm and found that the correlated insulating phases were killed when the screening gate was close by, but the superconductivity survived. Similarly, Ref.~\cite{Xiaoxue_Science2021_BLG} used a nearby Bernal-stacked bilayer graphene with varying carrier density to provide external screening. They also observed that superconductivity was more robust when Coulomb interaction was weakened by screening, and the critical temperature remained roughly constant over a wide range of doping in the screening layer. One might naively expect that these experiments rule out an electronic mechanism for superconductivity in twisted bilayer graphene. To the contrary, in this theoretical work, we show that the critical superconducting temperature predicted using the plasmon pairing mechanism is unchanged in the regimes probed in these experiments. For both experimental configurations, we find that the critical temperature is only suppressed for $l\lesssim3\mathrm{nm}$, which was not the case in either of the experiments. We can understand this as the length scale at which the bare Coulomb interaction is modified by the metal gates.

 In what follows we systematically investigate the robustness of the plasmon mechanism for superconductivity mediated by a screened Coulomb repulsion.  We first study the unscreened case, where it was previously shown~\cite{Girish_PRR2020_Plasmon} that the momentum-averaged Coulomb interaction can be modeled by a Lorentzian form.  In this limit, we prove that the critical temperature for the plasmonic mechanism is non-monotonic as a function of experimental parameters such as doping and twist angle.  We discuss this non-monotonicity in terms of averaged plasmon frequency and the unscreened Coulomb interaction.  For the screened Coulomb interaction, we find that the Lorentzian approximation no longer holds.  A full numerical calculation reveals that the critical temperature is not sensitive to the external screening gate unless it is closer than $l\approx 3\mathrm{nm}$. Finally, we solve the full-bandwidth Eliashberg equation and compute the tunneling density of state and find that, similar to the phonon-like mechanism, the plasmon mechanism gives a hard gap in the spectral function.

The paper is organized as follows. In Sec.~\ref{sec:Plasmonic_Superconductivity}, we introduce a minimal theoretical model for plasmonic superconductivity. In Sec.~\ref{sec:Plasmonic_SC_MATBG}, we discuss plasmon-induced superconductivity in magic angle twisted bilayer graphene. Sec.~\ref{sec:external_screening} focuses on superconductivity under external screening, including the hybrid double-layer structure and the metal gate structure. In Sec.~\ref{sec:STM_Spectrum}, we solve the full-bandwidth Eliashberg equation and compute the tunneling density of states. Finally, we discuss the conclusions and future directions of our work in Sec.~\ref{sec:Conclusions}. All derivations and technical details are provided in Appendices~\ref{appendix:Proven}-\ref{appendix:2lRPA}.

\section{Minimal model for plasmonic superconductivity} \label{sec:Plasmonic_Superconductivity}

We begin by introducing a minimal theoretical model that encapsulates the essential characteristics of plasmonic superconductivity. It has been shown that superconductivity mediated by purely electronic mechanism can be studied by introducing a momentum-averaged frequency-dependent Coulomb interaction~\cite{Grabowski_PRB1984_nonphonon, Girish_PRR2020_Plasmon}:
\begin{equation}
\lambda_{n,m}=N\left(E_F\right)\langle\langle V(i \omega_n-i \omega_m)\rangle\rangle,
\label{eq:coupling_strength}
\end{equation}
Here, $\lambda_{n,m}$ signifies the pairing strength of Cooper pairs, $N\left(E_F\right)$ represents the density of states (DOS) at the Fermi level, and $\langle\langle V(i \omega_n - i \omega_m)\rangle\rangle$ denotes the momentum-averaged Coulomb interaction:
\begin{equation}
\left\langle\left\langle V\left(i \omega_n\right)\right\rangle\right\rangle=\frac{\sum_{\mathbf{k}, \mathbf{p}} \Theta\left(\mathbf{k}_c-\mathbf{k}\right) \Theta\left(\mathbf{k}_c-\mathbf{p}\right) V\left(\mathbf{k}-\mathbf{p}, i \omega_n\right)}{\sum_{\mathbf{k}, \mathbf{p}} \Theta\left(\mathbf{k}_c-\mathbf{k}\right) \Theta\left(\mathbf{k}_c-\mathbf{p}\right)},
\label{average_coulomb}
\end{equation}
where $\mathbf{k}_c = 2\mathbf{k}_F$ represents the momentum cutoff, $\Theta$ is the Heaviside function, and $V\left(\mathbf{k}, i \omega_n\right)$ denotes the dynamically screened Coulomb interaction. Following the procedures proposed by Grabowski and Sham~\cite{Grabowski_PRB1984_nonphonon}, the linearized isotropic Eliashberg gap equation can be formulated as
\begin{equation}
\Delta_n=-2 \Tilde{T}_c \sum_{m=-\infty}^{\infty}  \frac{1}{Z_m \Tilde{\omega}_m} \arctan \frac{1}{Z_m  \Tilde{\omega}_m} \lambda_{n,m} \Delta_m,
\label{eq:iso_gap_eq}
\end{equation}
where $\widetilde{T}_c=k_B T_c/E_F$ represents the superconducting critical temperature, and $\widetilde{\omega}_n=(2 n+1) \pi \widetilde{T}_c$ denotes the dimensionless Matsubara frequency. Both quantities are scaled by the Fermi energy to be dimensionless. $Z_n$ is the mass renormalization function considering self-energy corrections, and $\Delta_n$ is the order parameter. Equation~(\ref{eq:iso_gap_eq}) has the form of an eigenvalue equation $\bar{\Delta} = \hat{C} \bar{\Delta}$, where the critical temperature $\widetilde{T}_c$ can be determined by identifying the largest eigenvalue of the matrix $\hat{C}$ that equals 1 (refer to Appendix~\ref{appendix:Linearized_ME_Theory} for details).

\begin{figure}[t]
    \includegraphics[width=0.8\columnwidth]{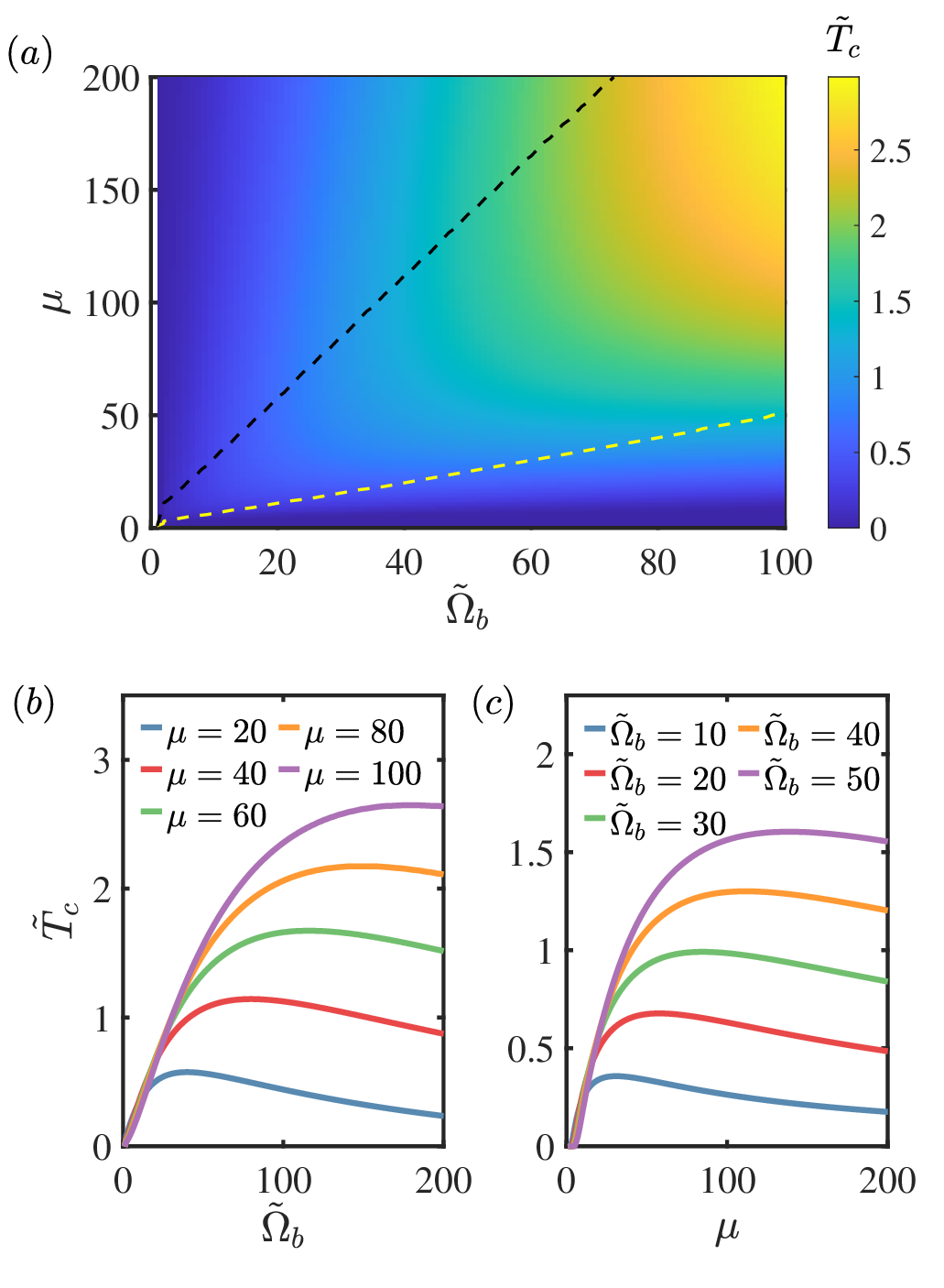}
    \caption{Critical temperature predicted by the Lorentzian model  (Eq.~(\ref{eq:Lorentzian})) for plasmon-mediated superconductivity valid in the absence of a screening gate.  Within this model, the critical temperature (scaled by Fermi energy) $\Tilde{T}_c$ is a function of two parameters: high-frequency limit of the Coulomb interaction $\mu$ and averaged (scaled) plasmon frequency $\Tilde{\Omega}_b$. (a) Color map of $\Tilde{T}_c(\mu, \Tilde{\Omega}_b) $.  The dashed black line traces the maximum critical temperature for fixed $\Tilde{\Omega}_b$, while the dashed yellow line traces the same for fixed $\mu$. (b) $\Tilde{T}_c$ versus $\Tilde{\Omega}_b$ for a given coupling strength $\mu$. (c) $\Tilde{T}_c$ versus $\mu$ for a given averaged (scaled) plasmon frequency $\Tilde{\Omega}_b$. In both cases, $\Tilde{T}_c$ shows non-monotonic behavior that can be understood analytically (see Appendix~\ref{appendix:Proven}).}
    \label{fig:tc_iso_gapeq}
\end{figure}

In general, $\lambda_{n,m}$ needs to be determined numerically due to the complexity of the Coulomb interaction. To simplify the problem, we adopt the Lorentzian approximation, which has been demonstrated to hold well in 2D systems~\cite{Grabowski_PRB1984_nonphonon}, allowing us to model the pairing interaction by the Lorentzian form:
\begin{equation}
\lambda_{n,m}=\mu\left(1- \frac{\Tilde{\Omega}_b^2}{\Tilde{\Omega}_b^2+(\Tilde{\omega}_n-\Tilde{\omega}_m)^2}\right),
\label{eq:Lorentzian}
\end{equation}
where $\mu$ represents the high-frequency limit of the pairing strength, set by both the bare Coulomb interaction and the density of states at the Fermi level. $\Tilde{\Omega}_b = \Omega_b/E_F$ represents the averaged (scaled) plasmon frequency, controlling the overall transition of the pairing strength from low to high frequency.

Figure~\ref{fig:tc_iso_gapeq}(a) shows the color map of the critical temperature $\Tilde{T}_c(\mu, \Tilde{\Omega}_b)$ within the Lorentzian approximation. The dashed black line traces the maximum critical temperature for a fixed $\Tilde{\Omega}_b$, while the dashed yellow line traces the same for a fixed $\mu$. Notably, we observe a non-monotonic behavior of the superconducting transition temperature for both fixed $\mu$ and $\Tilde{\Omega}_b$ cases in Fig.~\ref{fig:tc_iso_gapeq}(b) and Fig.~\ref{fig:tc_iso_gapeq}(c). In both cases, the non-monotonic behavior can be understood analytically, as detailed in Appendix~\ref{appendix:Proven}. A similar phenomenon has also been reported in recent works~\cite{Svistunov_PRX2023_Bipolaronic}, where non-monotonicity arises from the on-site repulsion $U$ in bond-Peierls bipolaronic superconductors. This unique non-monotonic behavior, absent in conventional phonon mechanisms, assumes significant importance in the context of MATBG, a point we will elucidate later.

\begin{figure}
    \includegraphics[width=0.75\columnwidth]{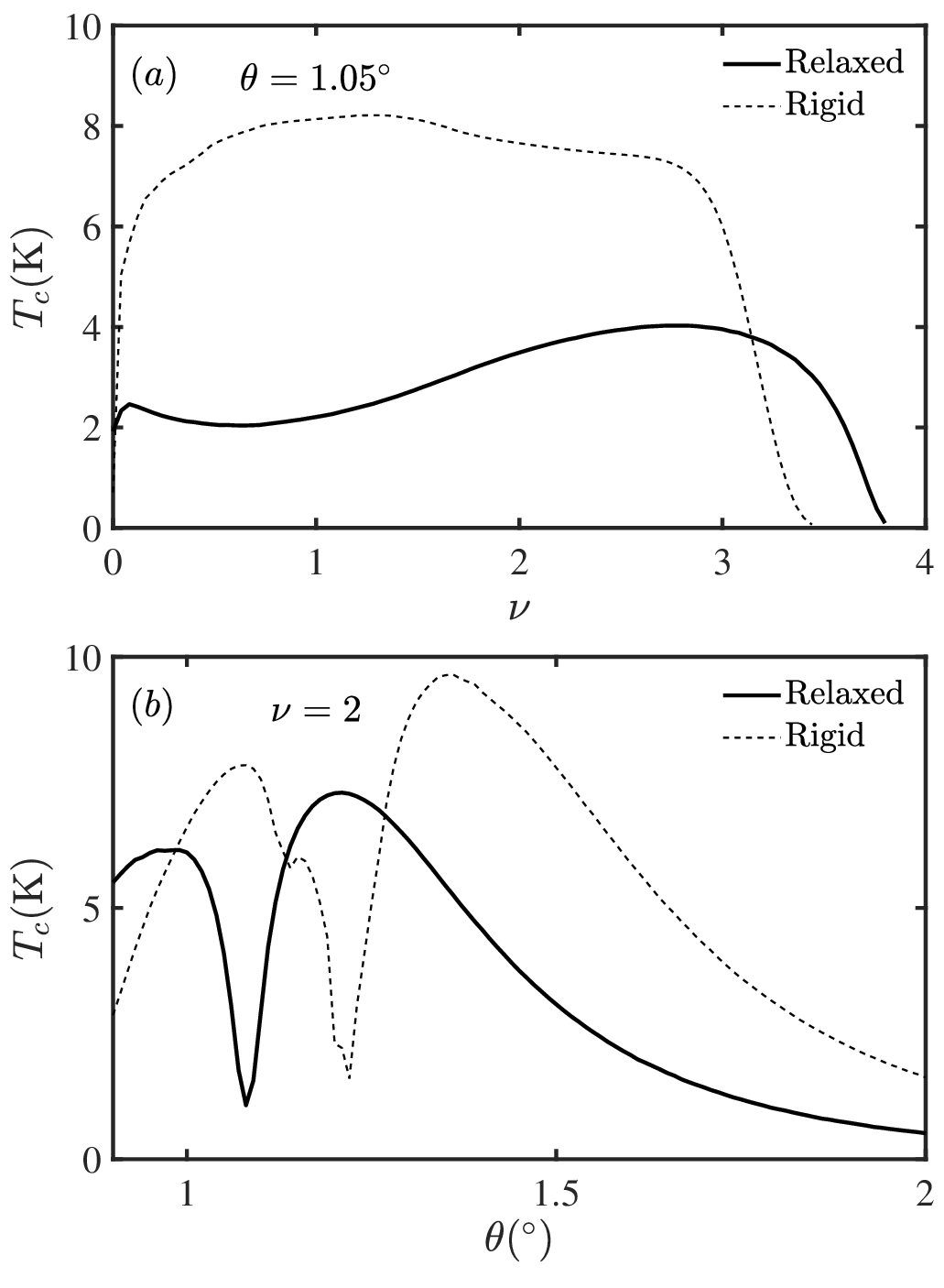}
    \caption{Plasmon-mediated superconducting critical temperature for magic-angle twisted bilayer graphene. (a) $T_c$ versus band filling factor both with and without lattice relaxation effects for $\theta = 1.05^\circ$.  The results show a weak bimodal feature.  The suppression of $T_c$ with relaxation is attributed partially to the change in magic angle when relaxation effects are included.  (b) $T_c$ versus twist angle for fixed filling factor $\nu=2$.  The bimodal dependence on twist angle persists even when lattice relaxation effects are included.}
    \label{fig:fitting_params}
\end{figure}

\begin{figure*}[t]
    \includegraphics[width=1.6\columnwidth]{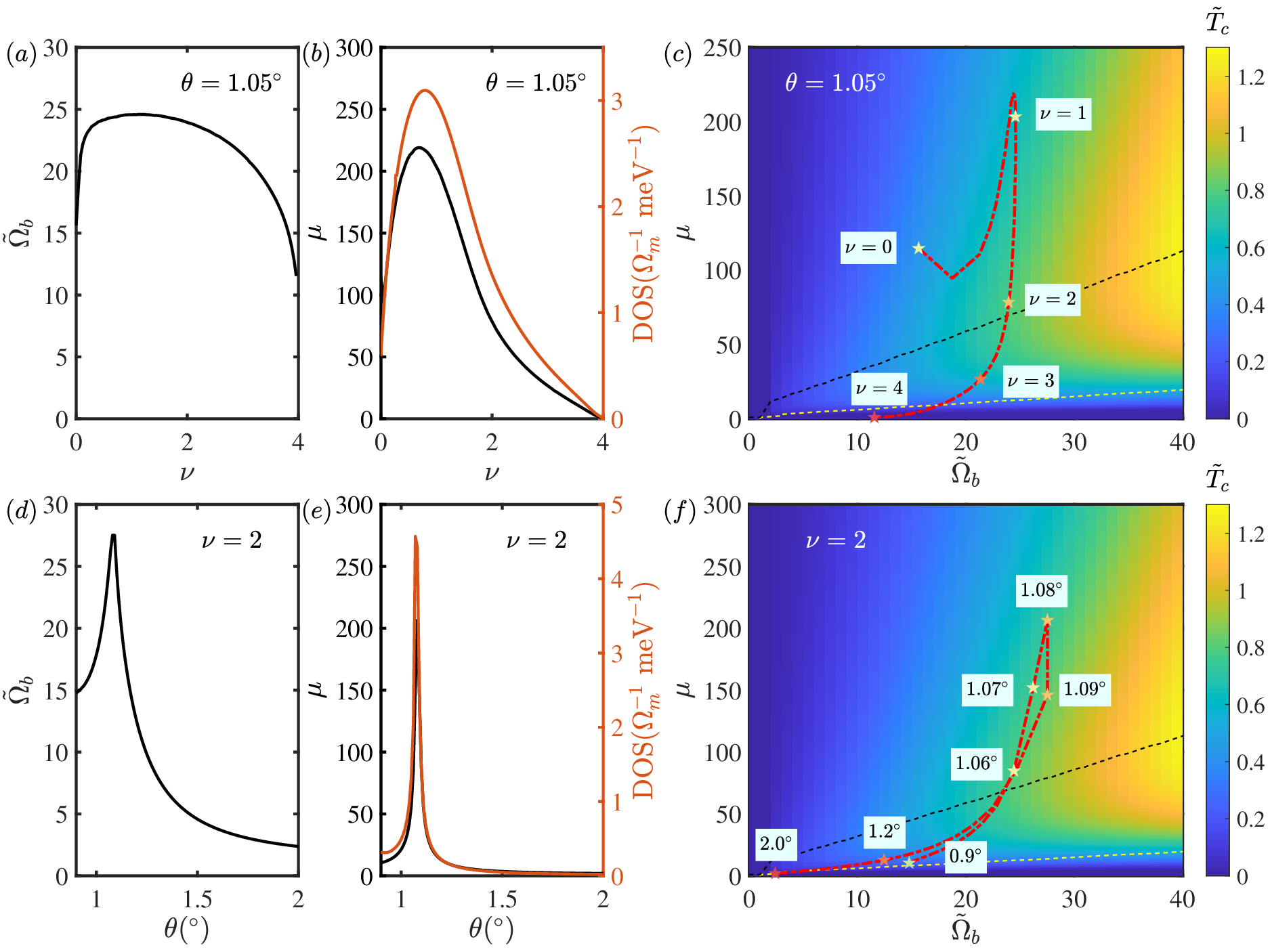}
    \caption{The non-monotonic behavior of critical temperature with twist-angle and filling can be understood by mapping the finite-temperature, finite-frequency random phase approximation calculation of the Bistritzer-MacDonald continuum model without a screening gate to the Lorentzian approximation used in this work. The averaged plasmon frequency depends weakly on filling factor (a) and strongly on twist angle (d).  By contrast, the high-frequency limit of the Coulomb interaction mostly just follows the density of states, shown in (b) and (e). The trajectory mapped in the $\Tilde{T}_c(\mu, \Tilde{\Omega}_b)$ phase space
    when changing filling is shown in (c), and similarly, shown in (f) for changing the twist angle.  In both cases, there is a non-monotonic dependence of $T_c$.  The maximum value of $\Tilde{T}_c$ is largely set by the momentum-averaged plasmon frequency, which within the BM continuum model does not exceed the values shown here.}
    \label{fig:tc_TBE_CM}
\end{figure*}

\section{Plasmonic superconductivity in magic angle twisted bilayer graphene} \label{sec:Plasmonic_SC_MATBG}

In this section, we will focus on superconductivity in MATBG. The dynamic properties of the screened Coulomb interactions for MATBG have been widely investigated in recent years~\cite{Levitov_PNAS2019_Intrinsically, Heinerich_NanoLett2016_plasmonic, Brey_RBL2020_Plasmonic, Marco_PRB2020_plasmonic, Shengjun_PRB2021_Collective, Mingwen_PRB2021_plasmon}. The intrinsically undamped plasmon modes were reported near magic angle which was believed to dominate in \moire system. In this work we consider the dynamically screened Coulomb interaction

\begin{equation}
V(\mathbf{q}, i \omega)=\frac{V(\mathbf{q})}{\epsilon(\mathbf{q}, i \omega)},
\label{eq:scr_coulomb_interaction}
\end{equation}
where $V(\mathbf{q})=2\pi e^2/\kappa q$ represents the bare Coulomb interaction, $\kappa$ is the background dielectric constant. We adopt $\kappa=3.03$ to account for the background dielectric subtraction for hexagonal boron nitride (hBN). $\epsilon(\mathbf{q}, i \omega)$ is the dynamic dielectric function calculated via random phase approximation (RPA)
\begin{equation}
\epsilon(\mathbf{q}, i \omega)=1-V(\mathbf{q})\Pi(\mathbf{q}, i \omega),
\label{eq:dielectric}
\end{equation}
and the polarization function $\Pi(\mathbf{q}, i \omega)$ is given by
\begin{equation}
\Pi(\mathbf{q}, i \omega)=2 \sum_{\mathbf{k}} \sum_{m, n} \frac{\left(f_{\mathbf{k}+\mathbf{q}}^n-f_{\mathbf{k}}^m\right) F_{\mathbf{k}, \mathbf{k}+\mathbf{q}}^{n m}}{E_{\mathbf{k}+\mathbf{q}}^n-E_{\mathbf{k}}^m-i \omega},
\label{polarization}
\end{equation}
where $m, n$ are band indices, $f_{\mathbf{k}}^m$ is the Fermi-Dirac distribution, and $F_{\mathbf{k}, \mathbf{k}+\mathbf{q}}^{n m}=\left|\psi_{n, \mathbf{k}+\mathbf{q}}^{\dagger} \psi_{m, \mathbf{k}}\right|^2$ is the form factor associated with different Bloch states. The additional factor $2$ accounts for the degeneracy of the spin and momentum $\mathbf{k}$ spans the whole Brillouin zone. Starting from the continuum model~\cite{MacDonald_PNAS2011_Moireband}, we apply RPA calculations to obtain the dynamic polarizability based on rigid/relaxed band structures, enabling us to compute the dynamic Coulomb interaction and momentum-averaged Coulomb interaction. For the realistic model, the Lorentzian approximation may not be valid, therefore we sought the full numerical solution instead (see Appendix~\ref{appendix:Linearized_ME_Theory} for more details).

Figure~\ref{fig:fitting_params}(a) shows the numerical results for the plasmon-mediated superconducting critical temperature in MATBG. We found that the lattice relaxation suppresses the superconductivity across a wide range of filling factors~\footnote{The band filling factor is defined as the ratio of the number of electrons required to fill the bands, which connect to the electronic density by $n_e=\nu/\Omega_m\approx\nu\times2.56\times10^{12}\mathrm{cm}^{-2}$ at $\theta=1.05^{\circ}$, where $n_e$ and $\Omega_m$ represent the electron density measured from charge natural point and area of \moire supercell.}. A weak bimodal feature is observed in the relaxed structures, which is similar to our previous observations~\cite{Girish_PRR2020_Plasmon}. This peak behavior qualitatively agrees with the filling factor window where the angle-dependent dome feature is reported in experiment. We also investigate the twist angle dependence as shown in Fig.~\ref{fig:fitting_params}(b). The bimodal structure persists even when the lattice relaxation effects are included. Relaxation acts as a redefinition of the magic angle, with a weak suppression of the critical temperature.

The non-monotonic behavior of the critical temperature with twist-angle and filling can be understood through the Lorentzian approximation discussed in Sec.~\ref{sec:Plasmonic_Superconductivity}. By fitting the frequency-dependent Coulomb interaction to the Lorentzian form (Eq~(\ref{eq:Lorentzian})), we can extract the corresponding parameters, as shown in Fig.~\ref{fig:tc_TBE_CM}. The results show that the averaged plasmon frequency $\Tilde{\Omega}_b$ depends weakly on filling but strongly on the twist angle, while the high-frequency limit of the Coulomb interaction $\mu$ mostly follows the density of states. The trajectory mapped in the $\Tilde{T}_c(\mu, \Tilde{\Omega}_b)$ phase space when changing the filling is shown in Fig.~\ref{fig:tc_TBE_CM}(c), and similarly, it is shown in Fig.~\ref{fig:tc_TBE_CM}(f) for changing the twist angle. In both cases, we observe a non-monotonic dependence of $T_c$. This non-monotonic behavior primarily results from the increase in $\mu$, which is due to the enhanced density of states near the magic angle. Additionally, it's worth noting that the maximum value of $\Tilde{T}_c$ is largely set by the momentum-averaged plasmon frequency, which within the BM continuum model does not exceed the values.

\section{The role of external screening} \label{sec:external_screening}

Now we turn to the case where external screening is involved. We start with the derivation of effective Coulomb interaction and then focus on hybrid double-layer structure and metal gate structure.

\subsection{Effective Coulomb interaction} \label{sec:external_scrrning_setup}

\begin{figure}[t]
    \includegraphics[width=0.85\columnwidth]{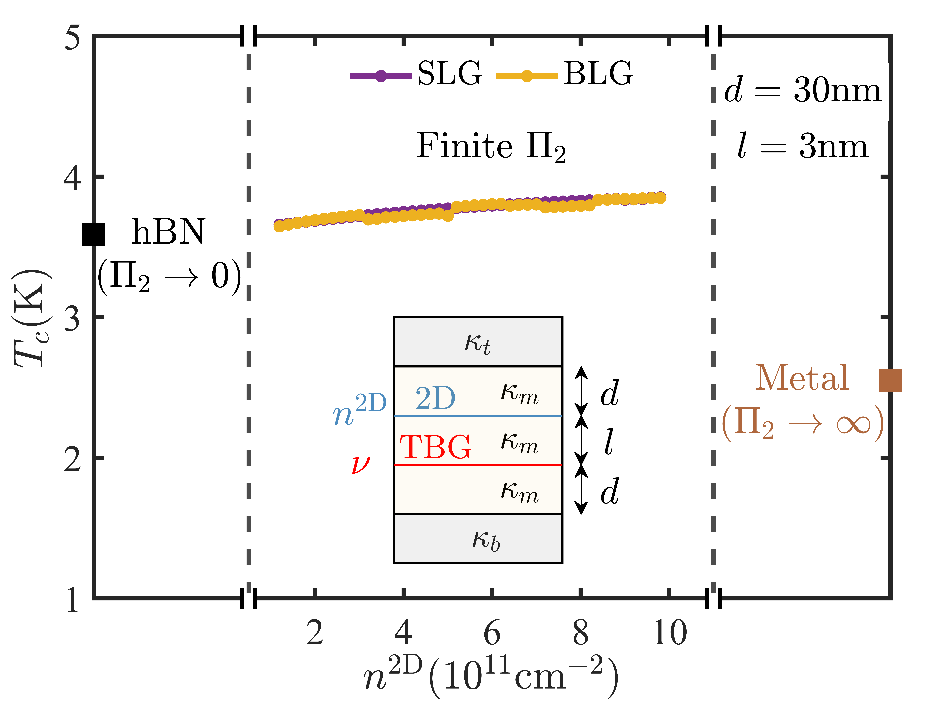}
    \caption{Critical temperature $T_c$ as a function of carrier density in screening layer $n^{\mathrm{2D}}$ for different screening materials. Here $\theta=1.05^{\circ}$ and $\nu=2$ were used for the MATBG layer. Purple and yellow lines represent the 2D material layer -- either single layer graphene (SLG) or bilayer graphene (BLG). Within the range of carrier density doping commonly used in experiments, the critical temperature is almost unchanged. The point labeled by ``hNB" and ``metal" represents the two limits discussed in the main text, corresponding to $\Pi_{2}\rightarrow0$ and $\Pi_{2}\rightarrow\infty$. The inset shows the schematic of the MATBG-2D heterostructure. }
    \label{fig:TBE_BLG}
\end{figure}

Inspired by the graphene double-layer experiments~\cite{PhysRevB.82.085443, PhysRevB.85.195444, pikalov2012graphene, doi:10.1063/1.4813821, Hosono_2014, kumar2020coulomb}, we consider the hybrid structure involving MATBG and a two-dimensional (2D) material, as depicted in the inset of Fig.~\ref{fig:TBE_BLG}. The structure consists of MATBG and the 2D material being separated by a distance $l$ within a homogeneous background with a dielectric constant $\kappa_m$. The MATBG layer is designated as the first layer, while the 2D layer constitutes the second layer. On either side of the background, materials with dielectric constants $\kappa_t$ and $\kappa_b$ are placed. For simplicity, we assume that both materials are at the same distance $d$ from the hybrid structure.

The electron-electron interaction among the charge carriers in MATBG is affected by the dielectric properties of the environment, which are encoded in the bare Coulomb interaction as introduced in Sec.~\ref{sec:Plasmonic_SC_MATBG}. When the second layer is introduced, the charge density in the second layer will alter the dielectric properties of the environment and the bare Coulomb interaction, thereby influencing its superconductivity. We analytically derive the effective bare Coulomb interaction and find 
\begin{equation}
V_{\mathrm{eff}}(\mathbf{q}, i \omega)=V_{11}(q)\left[1-\frac{V_{12}(q)V_{21}(q)}{V_{11}(q)V_{22}(q)}\left(1-\frac{1}{\epsilon_2(\mathbf{q}, i \omega)}\right)\right].
\label{eq:eff_bare_Coulomb}
\end{equation}
\noindent The detailed derivation of this result can be found in Appendix~\ref{appendix:2lRPA}.  Here, $\epsilon_2=1-V_{22}\Pi_{2}$ represents the dielectric function of the 2D layer.  Consequently, the dynamic screened Coulomb interaction within MATBG layer is 
\begin{equation}
V(\mathbf{q}, i \omega)=\frac{V_{\mathrm{eff}}(\mathbf{q}, i \omega)}{1-V_{\mathrm{eff}}(\mathbf{q}, i \omega)\Pi_1(\mathbf{q}, i \omega) }.\\
\end{equation}

Notably, Eq.~(\ref{eq:eff_bare_Coulomb}) enables us to determine the role of external screening on the superconducting properties of MATBG.  In particular, we can calculate how the dielectric properties of the screening layer $\epsilon_2(\mathbf{q}, i \omega)$ impact the Coulomb interaction within the MATBG layer.  We have checked that in the static limit, our result reproduces the effective screened Coulomb interaction previously used in the literature~\cite{Xiaoxue_Science2021_BLG}.

We first discuss two limiting cases, (i) where the screening layer is an insulator like bulk h-BN commonly used as a spacer in heterostructures, and (ii) where it is a metal like graphite or silicon used as gates in such structures.  For simplicity, we assume $\kappa_t=\kappa_b=\kappa_m=\kappa$, but this could be relaxed without any complication.  In the insulator limit, the screening layer is replaced by a background insulator, leading to a zero polarizability ($\Pi_{2}\rightarrow0$). The effective bare Coulomb interaction simplifies to 
\begin{equation}
\lim_{\Pi_{2}\rightarrow0} V_{\mathrm{eff}}(\mathbf{q}, i \omega)=V_{11}(q)=\frac{2\pi e^2}{\kappa q},
\end{equation}
which resembles the commonly used Coulomb interaction for 2D materials in a dielectric environment.  In the metallic limit, on the other hand, $\Pi_{2}\rightarrow\infty$. This results in 
\begin{equation}
\begin{aligned}
    \lim_{\Pi_{2}\rightarrow\infty} V_{\mathrm{eff}}(\mathbf{q}, i \omega)
    &=V_{11}(q)\left(1-\frac{V_{12}(q)V_{21}(q)}{V_{11}(q)V_{22}(q)}\right)\\
    &=\frac{2\pi e^2}{\kappa q}\left(1-e^{-2lq}\right),
\end{aligned}
\label{eq:metal_limit}
\end{equation}
which again corresponds to the standard case where MATBG is screened by an external metal~\cite{Lewandowski_PRB2021_umklapp, Reza_Plasmonics2020_Plasmon, Alonso_Natruenano2017_plasmons}.  These two limits serve to benchmark the role of external screening materials.  We note that the metallic limit is a lower bound for critical temperature, and in the presence of a screening layer $T_c$ should exceed this value.

\subsection{Numerical results} \label{sec:screening_layer}

To make a comparison with experiments, we mainly focus on two types of structures: (i) the hybrid double-layer structure~\cite{Xiaoxue_Science2021_BLG}, where the screening layer is a 2D semiconductor such as single-layer graphene (SLG) or bilayer graphene (BLG), and (ii) the metal gate structure~\cite{Levitov_Natrue2020_Untying, Saito_NatruePhys2020_TBG}, where a metal like graphite or silicon is used as gates.

In the first structure, we assume that the top and bottom layers are metals, i.e., $\kappa_t,\kappa_b\rightarrow\infty$. Figure~\ref{fig:TBE_BLG} shows the critical temperature $T_c$ as a function of the carrier density in the screening layer, denoted as $n^{\mathrm{2D}}$, for different screening materials. Here, we assume that the 2D material layer is either SLG (purple line) or BLG (yellow line). Within the range of carrier densities commonly used in experiments, the critical temperature remains almost unchanged. Our results align with those of Ref.~\cite{Xiaoxue_Science2021_BLG}, where a nearly constant critical temperature is observed in this structure. We also observe that the critical temperature is slightly higher than the insulator limit (the point labeled as ``hNB"), indicating that the superconductivity is stabilized by the screening layer. This finding is akin to an earlier experimental study~\cite{Arora_Natrue2020_stabilizedbyWSe2}, where insulating tungsten diselenide (WSe2) monolayers sandwiched between hBN and TBG contributed to the stabilization of superconductivity.

\begin{figure}
    \includegraphics[width=1\columnwidth]{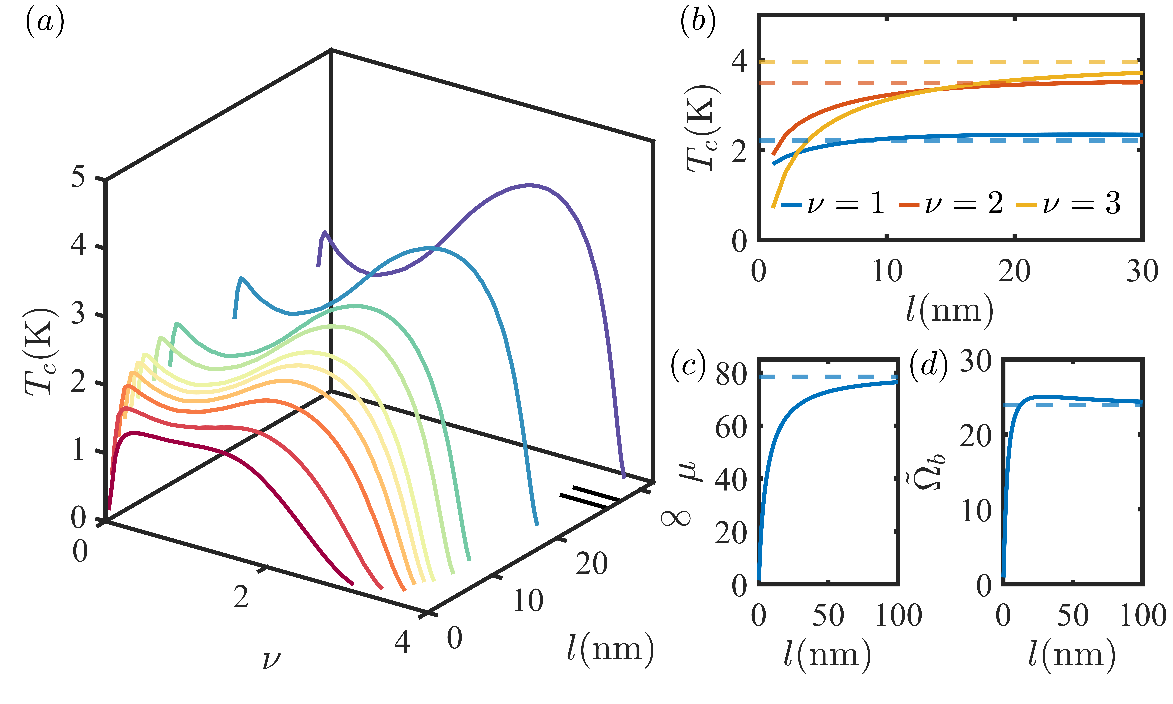}
    \caption{Numerical results for superconducting critical temperature with a single screening gate. (a) Critical temperature as a function of filling factor $\nu$ for different separation distances $l$. As $l$ decreases, the bimodal structure disappears and is replaced by a single peak near Van Hove singularity (VHS). (b) Critical temperature versus separation distance $l$ at fixed filling.  The dashed lines show the value for an infinite gate separation. (c) and (d) show the extracted parameters $\mu$ and $\Tilde{\Omega}_b$ for the Lorenzian model at $\nu=2$, where the dashed lines here show the value in the static screening limit. All calculations were performed with $\theta=1.05^{\circ}$. }
    \label{fig:tc_ext_gate_ssg}
\end{figure}

In the second structure, a single metal gate is considered as the screening layer, i.e., $\Pi_{2}\rightarrow\infty$. The corresponding bare Coulomb interaction has been discussed in Eq.~(\ref{eq:metal_limit}). Figure~\ref{fig:tc_ext_gate_ssg}(a) shows the critical temperature as a function of the filling factor $\nu$ for various separation distances $l$. As $l$ decreases, the bimodal feature disappears and is replaced by a single peak near the VHS. In Fig.~\ref{fig:tc_ext_gate_ssg}(b), a non-monotonic transition temperature is also observed for fixed filling factor cases. This non-monotonicity can also be understood by mapping to the Lorentzian model, as shown in Fig.~\ref{fig:tc_ext_gate_ssg}(c) and (d). Notably, we observed that the critical temperature remains relatively constant across a wide range of separations but experiences a significant drop when $l\lesssim 3 \mathrm{nm}$. However, the typical value of $l$ in experiments is around $7\mathrm{~nm}$ to $68\mathrm{~nm}$, which is larger than the distance where superconductivity is visibly suppressed.

The conclusion above can be understood as a result of the comparison between the size of the \moire Brillouin zone and the separation distance. According to Eq.~(\ref{eq:metal_limit}), the external metal gate will suppress the bare Coulomb interaction with a factor of $1 - e^{-2lq}$. The corresponding dynamic screened Coulomb interaction can be obtained by combining Eq.~(\ref{eq:scr_coulomb_interaction}) and Eq.~(\ref{eq:dielectric}), 
\begin{equation}
\begin{aligned}
    V(\mathbf{q}, i \omega)&=\frac{V_{0}(\mathbf{q})\left(1-e^{-2lq}\right)}{1-V_{0}(\mathbf{q})\left(1-e^{-2lq}\right)\Pi(\mathbf{q}, i \omega)}\\
    &=\frac{V_{0}(\mathbf{q})}{\frac{1}{1-e^{-2lq}}-V_{0}(\mathbf{q})\Pi(\mathbf{q}, i \omega)},\\
\end{aligned}
\label{eq:scr_coulomb_sggate}
\end{equation}
where $V_{0}(\mathbf{q})=2\pi e^2/\kappa q$ represents the bare Coulomb interaction without external screening. Eq.~(\ref{eq:scr_coulomb_sggate}) approaches zero in the limit of $l\rightarrow0$, indicating the Coulomb interaction is screened out when the external metal gate is close by. The dimensionless quantity $1/\left(1-e^{-2lq}\right)$ sets the scale for when the 
screening due to external metal gate is important. We note that due to the large lattice constant in \moire systems, a small separation distance is needed to screen out the Coulomb interaction. As approximation, we take the Fermi momentum $\mathbf{q}_F\approx q(\Gamma\rightarrow K)/2=2\pi/\sqrt{3}L_M/2\approx0.15\mathrm{~nm}^{-1}$, which gives $1/\left(1-e^{-2lq}\right) \approx 1.05$ for $l=10\mathrm{~nm}$, $1.29$ for $l=5\mathrm{~nm}$, and $1.69$ for $l=3\mathrm{~nm}$, indicating significant suppression happens only when $l\lesssim 3 \mathrm{nm}$.  Since $\ell/L_M$ is the relevant quantity, for even smaller twist angles (with larger $L_M$), this suppression could be achieved gate distances comparable to existing experiments.  Alternatively, we can understand this conclusion in terms of the superconducting coherence length that is known to be small in MATBG~\cite{Park_Nature2021_coherencelengthTTG, Davydov_PM2021_coherencelengthTBG}. According to BCS theory, the superconductivity coherence length is given by $\xi_0 = \hbar v_F/\pi \Delta_0$, where $v_F$ is the Fermi velocity and $\Delta_0$ is the quasi-particle gap. Approximating $\hbar v_F \approx 32 \mathrm{meV\cdot \AA}$ near the magic angle and $\Delta_0 \approx 2 \mathrm{meV}$ from tunneling density of states results (see Sec.~\ref{sec:STM_Spectrum}), we obtain $\xi_0 \approx 5 \mathrm{\AA}$. This small coherence length confirms that a small gate separation is necessary to disrupt the superconducting phase.

\section{Spectral Function and Tunneling Density of States} \label{sec:STM_Spectrum}

The connection between the tunneling density of states and scanning tunneling spectroscopy/microscopy (STS/STM) spectra has been understood for a long time~\cite{Thomas_PRL1963_STM1, Thomas_PRL1963_STM2}. Advances in angle-resolved photoemission spectroscopy (ARPES) have made it possible to compare theory with experiments~\cite{Vekhter_PRL2003_Proposal, Shi_PRL2004_Direct, Cuk_PSS2004_ARPES}. Recently, a series of experiments have attempted to measure the superconductivity gap in \moire systems~\cite{oh2021evidence, Kim_Nature2022_SCTTG}. In particular,  Ref.~\cite{oh2021evidence} reported strong spectroscopic evidence of a ``V"-shaped gap, suggesting an unconventional pairing. While some previous work has discussed the pairing symmetry for phonon-based superconductivity~\cite{Schrodi_PRR2020_Prominent, Schrodi_PRB2021_Cascade}, it remains unclear what the pairing symmetry is for the plasmonic mechanism. In this section, we will explore the signatures in the tunneling density of states as probed in spectroscopy experiments for both mechanisms.

\begin{figure}[t]
    \includegraphics[width=0.8\columnwidth]{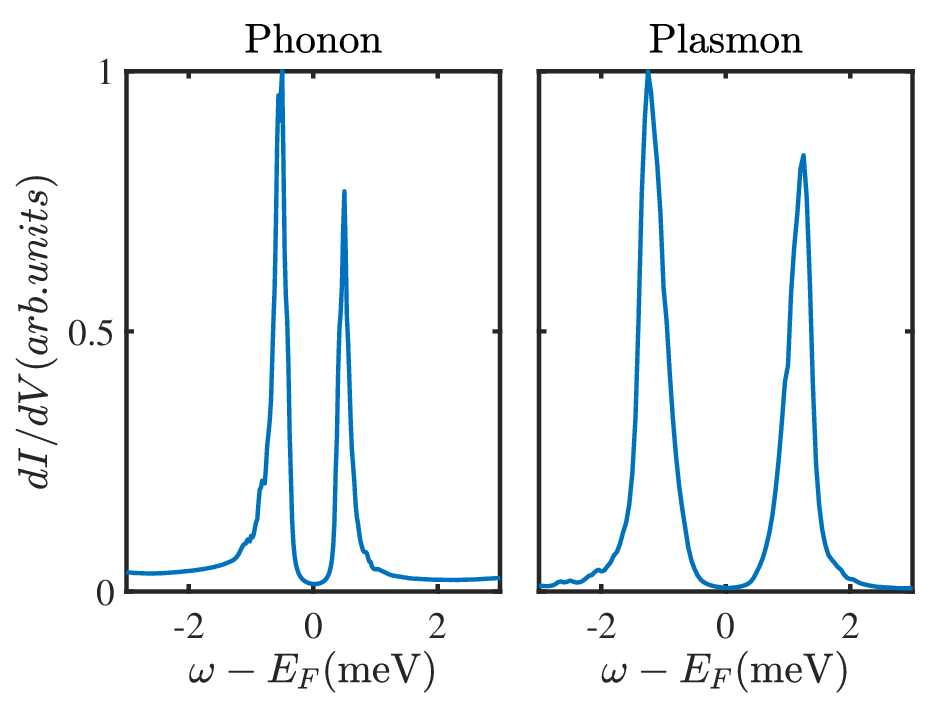}
    \caption{Tunneling density of state for both phonon and plasmon mechanism given by solving full-bandwidth Eliashberg equation.  Other than the asymmetry in the peaks, both phonons and plasmons show a relatively hard superconducting gap consistent with s-wave superconductivity.}
    \label{STM_Spectrum}
\end{figure}

To determine the tunneling density of states, it is necessary to obtain the superconductivity gap in real frequency, i.e., $i\omega_n\rightarrow\omega+i\eta$. This process can be achieved by employing the approach introduced by Marsiglio et al.~\cite{PhysRevB.37.4965}, where we need to solve the full-bandwidth Eliashberg equations in both the real and imaginary axes. This method provides an exact solution, but it is computationally intensive. To reduce computational demands, we first solve the full-bandwidth gap equations in the imaginary axis and then adopt the Pad\'{e} approximation~\cite{vidberg1977solving, leavens1985extension} to analytically continue the results to real frequencies. The anisotropic Eliashberg gap equations consist of three components: the order parameter $\phi$, the mass renormalization $Z$, and the chemical potential shift $\chi$ (see Appendix \ref{appendix:ME_Theory} for more details).  To simplify the calculations, we consider only the  singlet superconducting channel in our calculations and set $\chi=0$.  This simplification allows us to compare the spectral function from both the phonon-pairing and plasmon-pairing on equal footing.  After we obtain $Z(\mathbf{k},\omega)$ and $\phi(\mathbf{k},\omega)$, the band and momentum resolved spectral function can be computed via
\begin{equation}
A_n(\mathbf{k}, \omega)=-\frac{1}{\pi} \operatorname{Im}\left(\left[\hat{G}_{n}(\mathbf{k}, \omega+i \delta)\right]_{11}\right),
\end{equation}
where $n$ is the band index. The tunneling density of state is then computed by further summation over the momentum degree of freedom~\cite{Schrodi_PRR2020_Prominent, Schrodi_PRB2021_Cascade},
\begin{equation}
\frac{d I}{d V} \propto A(\omega)=\sum_{\mathbf{k}, n} A_n(\mathbf{k}, \omega).
\end{equation}

Figure~\ref{STM_Spectrum} shows the tunneling density of state for both phonon and plasmon mechanisms. Other than the asymmetry in the peaks, both phonons and plasmons show a relatively hard superconducting gap consistent with s-wave superconductivity. It is worth noting that both mechanisms yield similar spectral features, making them potentially challenging to differentiate directly through STM measurements.  Moreover, 
our calculations do not account for scattering from impurities, which has been shown to potentially dampen the spectral measurement and consequently soften the superconductivity gap.  In general, we observe that the superconducting gap for the plasmonic mechanism is bigger than the phonon mechanism. This is partially because the electron-electron interaction is stronger than the electron-phonon interaction in the flat band system.  

\section{Conclusions} \label{sec:Conclusions}

In this work, we primarily focus on superconductivity in MATBG mediated by electron-electron interaction within the framework of Eliashberg theory.  We show that in the absence of an external screening layer, the plasmonic superconductivity is well-described by a Lorentzian model giving a non-monotonic dependence of the transition temperature on experimentally tunable parameters, such as doping and twist angle, consistent with some of the features observed experimentally.  With external screening, this approximation no longer works, and our computationally intensive calculation shows that critical temperature is insensitive to the external screening gate unless this gate is closer than $l \approx 3$~nm.  We qualitatively understand this result as equivalent to $\l \ll L_M$; implying that for a gate at fixed separation, its screening will become more visible at smaller twist angles.        

At present, we are unable to conclude definitively the microscopic nature of the superconducting pairing in MATBG.  The higher transition temperature and the dome feature in both angle and filling predicted by the plasmon mechanism are both favorable features when compared to experimental observations.  However, there is much that this mechanism also gets incorrect including a hard gap of the spectral function, and a second dome-like feature at low-angles that is robust to relaxation effects.  Theoretically, there are properties not included in the Eliashberg theory, for example, it has been argued that it's important to consider the superfluid weight as a criterion for determining the superconducting phase in the strongly correlated regime~\cite{ShouchengZhang_PRB1993_criteria, Randeria_PRX2019_BoundsSCTc}. Moreover, it has been shown that the geometry of the band structure may also contribute to the superfluid weight~\cite{Rossi_PRL2019_GeometricContributionTBG, Julku_PRB2020_Superfluidweight}.  This interplay between transition temperature and superfluid weight for the plasmon-mediated pairing mechanism is an interesting question that we leave for future work.

\begin{acknowledgments}
 It is a pleasure to thank Alexey Bergyudin and Gargee Sharma for helpful comments and for collaboration on related projects.  We acknowledge the financial support from the Singapore National Research Foundation Investigator Award (NRF-NRFI06-2020-0003).
\end{acknowledgments}

\appendix

\section{Proof of non-monotonicity for Eq.~(\ref{eq:iso_gap_eq})}\label{appendix:Proven}

In this section, we will prove that the solution of Eq.~(\ref{eq:iso_gap_eq}) is non-monotonic. We first noticed that the coupling strength in Eq.~(\ref{eq:coupling_strength}) becomes a constant value in the limit of (i) $\mu\rightarrow0$, (ii) $\Tilde{\Omega}_{b}\rightarrow0$, and (iii) $\Tilde{\Omega}_{b}\rightarrow\infty$. We assume $\lambda_{nm}=\lambda_0\ne0$ without loss of generality, which allows us to rewrite Eq.~(\ref{eq:iso_gap_eq}) as:
\begin{equation}
\Delta_n=-2 \Tilde{T}_c  \lambda_{0} \sum_{m=-\infty}^{\infty}  \frac{1}{Z_m \Tilde{\omega}_m} \arctan \frac{1}{Z_m  \Tilde{\omega}_m} \Delta_m.
\label{iso_gap_eq_limit}
\end{equation}

In those limits, Eq.~(\ref{iso_gap_eq_limit}) becomes a decoupled equation: the summation on right-hand side is independent of $n$, indicating that the order parameter on the left-hand side is constant. Considering the nontrivial solution $\Delta_n=\Delta_0\neq0$, $\Delta_n$ can be safely canceled from both sides:
\begin{equation}
1=-2 \Tilde{T}_c  \lambda_{0} \sum_{m=-\infty}^{\infty}  \frac{1}{Z_m \Tilde{\omega}_m} \arctan \frac{1}{Z_m  \Tilde{\omega}_m}.
\label{iso_gap_eq_limit1}
\end{equation}

However, the summation on right-hand always gives a negative value. To reconcile the signs on both sides, Eq.~(\ref{iso_gap_eq_limit1}) only allows the trivial solution $\Tilde{T}_c=0$. 

In the limit of (iv) $\mu\rightarrow\infty$, we obtain the limit below:
\begin{equation}
\begin{aligned}
\lim_{\mu\rightarrow\infty}\frac{1}{Z_n \Tilde{\omega}_{n}}&\rightarrow\frac{1}{\mu\frac{\Tilde{\Omega}_b}{\Tilde{\omega}_n} \arctan \frac{\Tilde{\omega}_n }{\Tilde{\omega}_n^2+\Tilde{\Omega}_b\left(1+\Tilde{\Omega}_b\right)}}\rightarrow0,\\
\lim_{\mu\rightarrow\infty}\frac{\mu}{Z_n \Tilde{\omega}_{n}}&\rightarrow\frac{1}{\frac{\Tilde{\Omega}_b}{\Tilde{\omega}_n} \arctan \frac{\Tilde{\omega}_n }{\Tilde{\omega}_n^2+\Tilde{\Omega}_b\left(1+\Tilde{\Omega}_b\right)}}>0,
\label{eq:muinf_expressions}
\end{aligned}
\end{equation}
where we have used the analytical solution of $Z_n$ derived in Eq.~(\ref{eq:ana_solution_Z}) (see also~\cite{Grabowski_PRB1984_nonphonon}).  By substituting Eq.~(\ref{eq:Lorentzian}) into Eq.~(\ref{eq:iso_gap_eq}), we obtain:
\begin{equation}
\begin{aligned}
\Delta_n&=-2 \Tilde{T}_c \\
\times \sum_{m=-\infty}^{\infty}  &\frac{\mu}{Z_m \Tilde{\omega}_m} \arctan \frac{1}{Z_m  \Tilde{\omega}_m} \left(1- \frac{\Tilde{\Omega}_b^2}{\Tilde{\Omega}_b^2+(\Tilde{\omega}_n-\Tilde{\omega}_m)^2}\right) \Delta_m.
\label{eq:iso_gap_eq_infmu}
\end{aligned}
\end{equation}

Using the expressions listed Eq.~(\ref{eq:muinf_expressions}), we find that $\lim_{\mu\rightarrow\infty}\arctan((Z_n \Tilde{\omega}_{n})^{-1})=0$. Consequently, the summation over index $m$ gives a zero contribution in Eq.~(\ref{eq:iso_gap_eq_infmu}), leading to a trivial solution, $\Tilde{T}_c=0$.

In summary, we have demonstrated that $\Tilde{T}_c$ approaches zero in the limits of $\mu\rightarrow0$ and $\mu\rightarrow\infty$, as well as $\Tilde{\Omega}_b\rightarrow0$ and $\Tilde{\Omega}_b\rightarrow\infty$. If there exists a non-zero solution for the critical temperature in $\Tilde{T}_c(\mu, \Tilde{\Omega}_b)$ space, the solution for the gap equation must exhibit non-monotonic behavior.

\section{Details of the continuum model}\label{appendix:continuum_model}

Here we provide a brief review of the continuum model introduced by Ref.~\cite{MacDonald_PNAS2011_Moireband}. We begin with AA-stacked bilayer graphene with rotation $-\theta/2$ and $\theta/2$ for layers 1 and 2. The lattice vector before rotation is defined as $\mathbf{a}_1=a(1,0)$ and $\mathbf{a}_2=a(1 / 2, \sqrt{3} / 2)$, where $a=2.46~\mathrm{\AA}$ is the lattice constant for monolayer graphene. The corresponding reciprocal lattice vectors are $\mathbf{a}_1^*=(2 \pi / a)(1,-1 / \sqrt{3})$ and $\mathbf{a}_2^*=(2 \pi / a)(0,2 / \sqrt{3})$. After we apply the rotation, the lattice vectors and reciprocal lattice vectors are given by $\mathbf{a}_i^{(l)}=R(\mp \theta / 2) \mathbf{a}_i$ and $\mathbf{a}_i^{*(l)}=R(\mp \theta / 2) \mathbf{a}_i^*$, with $\mp$ for $l=1,2$, respectively, where $R( \pm \theta / 2)$ is the rotation matrix. The Dirac points for rotated graphene are located at $\mathbf{K}_{\xi}^{(l)}=-\xi\left[2 \mathbf{a}_1^{(l) *}+\mathbf{a}_2^{(l) *}\right] / 3$ for layer $l$, where $\xi=\pm 1$ is the valley index.

In the case of a small twist angle, the commensurate structure can be approximately defined. The reciprocal lattice vectors for \moire Brillouin zone are given by $\mathbf{G}_i^{\mathrm{M}}=\mathbf{a}_i^{(1)}-\mathbf{a}_i^{(2)}$ for $i=1,2$. The effective Hamiltonian of the continuum model in valley $\xi$ takes the form
\begin{equation}
H^{(\xi)}=\left(\begin{array}{cc}
H_1 & U \\
U^{\dagger} & H_2
\end{array}\right),
\end{equation}
in the basis of $(A_1,B_1,A_2,B_2)$ site. Here $H_l$ is the intralayer Hamiltonian for layer $l$
\begin{equation}
H_l=-\hbar v\left[R( \pm \theta / 2)\left(\mathbf{k}-\mathbf{K}_{\xi}^{(l)}\right)\right] \cdot\left(\xi \sigma_x, \sigma_y\right),
\end{equation}
and $U$ is the interlayer coupling
\begin{equation}
\begin{aligned}
U= & \left(\begin{array}{cc}
w_{AA} & w_{AB} \\
w_{AB} & w_{AA}
\end{array}\right)+\left(\begin{array}{cc}
w_{AA} & w_{AB} \omega^{-\xi} \\
w_{AB} \omega^{\xi} & w_{AA}
\end{array}\right) e^{i \xi \mathbf{G}_1^{\mathrm{M}} \cdot \mathbf{r}} \\
& +\left(\begin{array}{cc}
w_{AA} & w_{AB} \omega^{\xi} \\
w_{AB} \omega^{-\xi} & w_{AA}
\end{array}\right) e^{i \xi\left(\mathbf{G}_1^{\mathrm{M}}+\mathbf{G}_2^{\mathrm{M}}\right) \cdot \mathbf{r}},
\end{aligned}
\end{equation}
where $\omega=e^{2i\pi/3}$ and $L_M=a/2\sin{(\theta/2)}$ is lattice constant for real space. In this paper, we use the parameters $\hbar v = 5250\ \mathrm{meV\cdot\AA}$, $w_{AA} = 79.7\ \mathrm{meV}$, and $w_{AB} = 97.5\ \mathrm{meV}$ due to the relaxation effect~\cite{Koshino_PRX2018_Wannier}. For a more detailed analysis of the origin of the Hamiltonian, we refer the reader to Ref.~\cite{Koshino_PRX2018_Wannier, Koshino_PRB2020_continuummodel}. For a given Bloch vector $\mathbf{k}$ in the moiré Brillouin zone, there are many states associated with each other by the interlayer coupling matrix $U$, which can be mapped by $\mathbf{q} = \mathbf{k} + \mathrm{n}\mathbf{G}_{1}^{M} + \mathrm{m}\mathbf{G}_{2}^{M}$, where $n$ and $m$ are integers. To ensure convergence, we select the state inside the circle $\left|\mathbf{q}-\mathrm{q}_0\right|<q_c$, where $\mathbf{q}_0$ is the midpoint between $\mathbf{K}_{\xi}^{(1)}$ and $\mathbf{K}_{\xi}^{(2)}$, and $q_c$ is set to $4 G_{\mathrm{M}}\left(G_{\mathrm{M}}=\left|\mathbf{G}_1^{\mathrm{M}}\right|=\left|\mathbf{G}_2^{\mathrm{M}}\right|\right)$. The calculation is done independently for each valley.

\section{Migdal-Eliashberg theory} \label{appendix:ME_Theory}

The Eliashebrg theory was built within the framework of the Nambu-Gor’kov formalism~\cite{PhysRev.117.648, eliashberg1960interactions, eliashberg1961temperature, PhysRevB.87.024505, MARSIGLIO2020168102, gor1959microscopic}. The two-component electron spinor within the formalism is written down as 
\begin{equation}
\psi_{\mathbf{k}}=\left(\begin{array}{c}
c_{\mathbf{k} \uparrow} \\
c_{-\mathbf{k} \downarrow}^{\dagger}
\end{array}\right), \quad \psi_{\mathbf{k}}^{\dagger}=\left(\begin{array}{cc}
c_{\mathbf{k} \uparrow}^{\dagger} & c_{-\mathbf{k} \downarrow}
\end{array}\right),
\end{equation}
where the operator $c_{\mathbf{k} \uparrow}$ ($c_{-\mathbf{k} \downarrow}^{\dagger}$) destroy (create) an electron of Bloch state in momentum $\mathbf{k}$ ($-\mathbf{k}$) and spin up (down). By this definition, the Green function of electron is $2\times2$ matrix
\begin{equation}
\hat{G}(\mathbf{k}, \tau)=-\left[\begin{array}{cc}
\left\langle \mathcal{T} c_{\mathbf{k} \uparrow}(\tau) c_{\mathbf{k} \uparrow}^{\dagger}(0)\right\rangle & \left\langle \mathcal{T} c_{\mathbf{k} \uparrow}(\tau) c_{-\mathbf{k} \downarrow}(0)\right\rangle \\
\left\langle \mathcal{T} c_{-\mathbf{k} \downarrow}^{\dagger}(\tau) c_{\mathbf{k} \uparrow}^{\dagger}(0)\right\rangle & \left\langle \mathcal{T} c_{-\mathbf{k} \downarrow}^{\dagger}(\tau) c_{-\mathbf{k} \downarrow}(0)\right\rangle
\end{array}\right],
\label{gl_Green}
\end{equation}
where $\mathcal{T}$ is the time-ordering operator and the braces indicate a grand-canonical thermodynamic average. Here the diagonal term is the conventional electron Green function and the off-diagonal terms represent Gor’kov’s anomalous Green functions $F(\mathbf{k}, \tau)$ and $F^{*}(\mathbf{k}, \tau)$, which describe the energy properties of superconducting state. The Green function can then be expanded using the Fourier series:
\begin{equation}
\hat{G}(\mathbf{k}, \tau)=T\sum_{i \omega_n} e^{-i \omega_n \tau} \hat{G}\left(\mathbf{k}, i \omega_n\right),
\label{fouier_tf}
\end{equation}
where $T$ is temperature and $\omega_n=(2 n+1) \pi T_C$ is the Matsubara frequency. Combining Eq.~(\ref{gl_Green}) and Eq.~(\ref{fouier_tf}), we obtain the Green function in momentum space in imaginary frequency
\begin{equation}
\hat{G}\left(\mathbf{k}, i \omega_n\right)=\left[\begin{array}{cc}
G\left(\mathbf{k}, i \omega_n\right) & F\left(\mathbf{k}, i \omega_n\right) \\
F^*\left(\mathbf{k}, i \omega_n\right) & -G\left(-\mathbf{k},-i \omega_n\right)
\end{array}\right].
\label{fouier_tf_im}
\end{equation}

The Eliashberg theory aims to solve the generalized Green function Eq.~(\ref{fouier_tf_im}) using Dyson equation
\begin{equation}
\hat{G}^{-1}\left(\mathbf{k}, i \omega_n\right)=\hat{G}_0^{-1}\left(\mathbf{k}, i \omega_n\right)-\hat{\Sigma}\left(\mathbf{k}, i \omega_n\right),
\label{dyson}
\end{equation}
where $\hat{G}_0^{-1}\left(\mathbf{k}, i \omega_n\right)$ is non-interacting Green function given by
\begin{equation}
\hat{G}_0^{-1}\left(\mathbf{k}, i \omega_n\right)=i \omega_n \hat{\tau}_0-\xi_{\mathbf{k}} \hat{\tau}_3,
\label{NI_green}
\end{equation}
where $\xi_{\mathbf{k}}=E_\mathbf{k}-E_{\mathrm{F}}$ and $\hat{\Sigma}\left(\mathbf{k}, i \omega_n\right)$ is self-energy. In general, it is very difficult to solve exact self-energy due to the complexity of the phonon propagator. However, Migdal’s theorem states that the phonon vertex corrections are small~\cite{migdal1958interaction}. It's therefore a good approximation to set the phonon vertex to the bare phonon vertex. Within the Migdal-Eliashberg approximation, one can write down self-energy as
\begin{equation}
\hat{\Sigma}\left(\mathbf{k}, i \omega_n\right)=  -T \sum_{\mathbf{k}^{\prime} m}  V_{\mathbf{k}-\mathbf{k}^{\prime}, n-m}
 \hat{\tau}_3 \hat{G}\left(\mathbf{k}^{\prime}, i \omega_{m}\right) \hat{\tau}_3.
\label{Migdal}
\end{equation}
The interaction $V_{\mathbf{k}-\mathbf{k}^{\prime}, n-m}$ defined as
\begin{equation}
\begin{aligned}
V_{\mathbf{k}-\mathbf{k}^{\prime}, n-m}&
= \{ \left|g_{\mathbf{k k}^{\prime}}\right|^2 D\left(\mathbf{k}-\mathbf{k}^{\prime}, i \omega_n-i \omega_{m}\right) \\
& -  V_c\left(\mathbf{k}-\mathbf{k}^{\prime}, i \omega_{n}-i\omega_{m}\right)\} \left|M(\mathbf{k},\mathbf{k}^{\prime})\right|^2,
\end{aligned}
\end{equation}
where $D\left(\mathbf{q}, i \omega_n\right)=2\omega_{\mathbf{q}}/\left[ (i\omega_n)^2-\omega_{\mathbf{q}}^2\right]$ is the dressed phonon propagator in momentum $\mathbf{q}$, 
$\left|g_{\mathbf{k k}^{\prime}}\right|^2=\hbar D^2 q/2A\rho c_{ph}$ is the electron-phonon coupling. $\omega_{\mathbf{q}}=c_{ph}q$ is phonon dispersion, $c_{ph}$ is the phonon velocity, $D$ is the deformation potential, $A$ is sample area, $\rho$ is the mass density, and $|\psi_{\mathbf{k}}\rangle$ is the wavefunction for momentum $\mathbf{k}$. $M(\mathbf{k},\mathbf{k}^{\prime})=\langle\psi_{\mathbf{k}}|\psi_{\mathbf{k}^{\prime}}\rangle$ is the form factor which involved the projection to the occupied bands~\cite{Patrick_PRB2017_bismuth, Bernevig_PRB2021_TBG}. In this paper, we use $D=25\ \mathrm{eV}, \ c_{ph}=20000\ \mathrm{m/s},\ \rho=7.6\times10^{-8}\ \mathrm{g/cm^2}$~\cite{PhysRevLett.105.256805, chen2008intrinsic}. 

Assuming the ansatz for self-energy
\begin{equation}
\begin{aligned}
\hat{\Sigma}\left(\mathbf{k}, i \omega_n\right)= & i \omega_n\left[1-Z_{\mathbf{k}, n}\right] \hat{\tau}_0+\chi_{\mathbf{k}, n} \hat{\tau}_3 \\
& +\phi_{\mathbf{k}, n} \hat{\tau}_1+\bar{\phi}_{\mathbf{k}, n} \hat{\tau}_2,
\end{aligned}
\label{ansatz}
\end{equation}
where $Z_{\mathbf{k}, n}$ is the mass renormalization function, $\chi_{\mathbf{k}, n}$ is the chemical potential shift, and $\phi_{\mathbf{k}, n}$ is the order parameter. If the phase of the superconductivity gap is not important, one can choose the gauge $\bar{\phi}_{\mathbf{k}, n}=0$. Combining Eq.~(\ref{dyson}), Eq.~(\ref{NI_green}), Eq.~(\ref{Migdal}) and Eq.~(\ref{ansatz}), we arrive at the anisotropic Eliashberg equation
\begin{equation}
\begin{aligned}
i\omega_n\left(1-Z_{\mathbf{k}, n}\right)&=T \sum_{\mathbf{k}^{\prime} m}  V_{\mathbf{k}-\mathbf{k}^{\prime}, n-m}\frac{i\omega_{m}Z_{\mathbf{k^{\prime}}, m}}{\Theta_{\mathbf{k^{\prime}}, m}} \\
\chi_{\mathbf{k}, n} &=T \sum_{\mathbf{k}^{\prime} m} V_{\mathbf{k}-\mathbf{k}^{\prime}, n-m}\frac{\xi_{\mathbf{k}}+\chi_{\mathbf{k^{\prime}}, m}}{\Theta_{\mathbf{k^{\prime}}, m}}\\
\phi_{\mathbf{k}, n} &=-T \sum_{\mathbf{k}^{\prime} m} V_{\mathbf{k}-\mathbf{k}^{\prime}, n-m}  \frac{\phi_{\mathbf{k^{\prime}}, m}}{\Theta_{\mathbf{k^{\prime}}, m}}, \\
\end{aligned}
\end{equation}
where $\Theta$ is the denominator defined as
\begin{equation}
\Theta_{\mathbf{k}, n}= {\left(\omega_n Z_{\mathbf{k}, n}\right)^2+\left(\xi_{\mathbf{k}}+\chi_{\mathbf{k}, n}\right)^2 } + \phi_{\mathbf{k}, n}^2.
\label{denominator}
\end{equation}

\section{Linearized Isotropic Gap Equation}\label{appendix:Linearized_ME_Theory}

In order to determine the critical temperature, we consider the linearized Eliashberg equation by setting $\phi\left(\mathbf{k}, i \omega_n\right)=0$ in the denominator of Eq.~(\ref{denominator}) and we have
\begin{equation}
\begin{aligned}
i\omega_n\left(1-Z_{\mathbf{k}, n}\right)&=T \sum_{\mathbf{k}^{\prime} m}  V_{\mathbf{k}-\mathbf{k}^{\prime}, n-m}\frac{i\omega_{m}Z_{\mathbf{k^{\prime}}, m}}{\omega_n^2 Z^2_{\mathbf{k}^{\prime}, m}+\left(\xi_{\mathbf{k}^{\prime}}+\chi_{\mathbf{k}^{\prime}, m}\right)^2} \\
\chi_{\mathbf{k}, n} &=T \sum_{\mathbf{k}^{\prime} m} V_{\mathbf{k}-\mathbf{k}^{\prime}, n-m}\frac{\xi_{\mathbf{k}}+\chi_{\mathbf{k^{\prime}}, m}}{\omega_n^2 Z^2_{\mathbf{k}^{\prime}, m}+\left(\xi_{\mathbf{k}^{\prime}}+\chi_{\mathbf{k}^{\prime}, m}\right)^2}\\
\phi_{\mathbf{k}, n} &=-T \sum_{\mathbf{k}^{\prime} m} V_{\mathbf{k}-\mathbf{k}^{\prime}, n-m}  \frac{\phi_{\mathbf{k^{\prime}}, m}}{\omega_n^2 Z^2_{\mathbf{k}^{\prime}, m}+\left(\xi_{\mathbf{k}^{\prime}}+\chi_{\mathbf{k}^{\prime}, m}\right)^2}. \\
\end{aligned}
\label{l_gap_eq}
\end{equation}

Combining first two equations of Eq.~(\ref{l_gap_eq}), we get
\begin{equation}
\begin{aligned}
R_{\mathbf{k}, n} &= i\omega_n\left(1-Z_{\mathbf{k}, n}\right)+\chi_{\mathbf{k}, n} \\
&=T \sum_{\mathbf{k}^{\prime} m}  V_{\mathbf{k}-\mathbf{k}^{\prime}, n-m}\frac{i\omega_{m}Z_{\mathbf{k^{\prime}}, m}+\xi_{\mathbf{k}}+\chi_{\mathbf{k^{\prime}}, m}}{\omega_n^2 Z^2_{\mathbf{k}^{\prime}, m}+\left(\xi_{\mathbf{k}^{\prime}}+\chi_{\mathbf{k}^{\prime}, m}\right)^2} \\
&=-T \sum_{\mathbf{k}^{\prime} m}  \frac{V_{\mathbf{k}-\mathbf{k}^{\prime}, n-m}}{i\omega_m Z_{\mathbf{k}^{\prime} m}-\xi_{\mathbf{k}^{\prime}}-\chi_{\mathbf{k}^{\prime}, m}} \\
&=-T \sum_{\mathbf{k}^{\prime} m}  \frac{V_{\mathbf{k}-\mathbf{k}^{\prime}, n-m}}{i\omega_m-\xi_{\mathbf{k}^{\prime}}-\left( i\omega_m\left(1-Z_{\mathbf{k}^{\prime}, n}\right)+\chi_{\mathbf{k}^{\prime}, m} \right)} \\
&=-T \sum_{\mathbf{k}^{\prime} m}  \frac{V_{\mathbf{k}-\mathbf{k}^{\prime}, n-m}}{i\omega_m-\xi_{\mathbf{k}^{\prime}}-R_{\mathbf{k}^{\prime}, m}}, \\
\end{aligned}
\label{reduce_self_energy}
\end{equation}
thus Eq.~(\ref{reduce_self_energy}) is a self-consistent equation containing all information about mass renormalization and chemical potential shift. The first-order self-energy correction can now be evaluated by setting $R_{\mathbf{k}^{\prime}, m}=0$ in right hand side of Eq.~(\ref{reduce_self_energy}),
\begin{equation}
\begin{aligned}
R_{\mathbf{k}, n}
&=-T_c \int_{-1}^{1}d \Tilde{E} \sum_{m}\frac{\lambda_{n,m}}{i\Tilde{\omega}_m-\Tilde{E}}. \\
\end{aligned}
\label{1st_self_corr}
\end{equation}

\begin{figure}[t]
    \includegraphics[width=1\columnwidth]{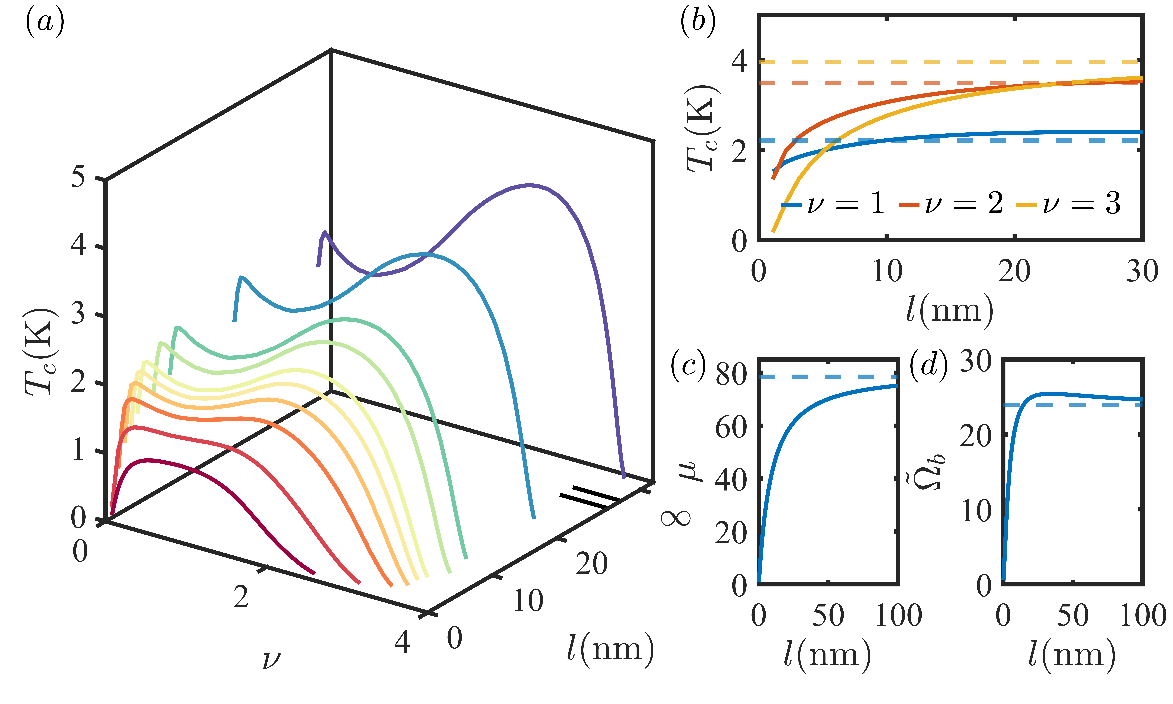}
    \caption{Numerical results for superconducting critical temperature with a double screening gate. (a) Critical temperature as a function of filling factor $\nu$ for different separation distances $l$. As $l$ decreases, the bimodal structure disappears and is replaced by a single peak near Van Hove singularity (VHS). (b) Critical temperature versus separation distance $l$ at fixed filling.  The dashed lines show the value for an infinite gate separation. (c) and (d) show the extracted parameters $\mu$ and $\Tilde{\Omega}_b$ for the Lorenzian model, where the dashed lines here show the value in the static screening limit.}
    \label{fig:tc_ext_gate_sdb}
\end{figure}

Follow paper~\cite{Grabowski_PRB1984_nonphonon, Girish_PRR2020_Plasmon}, we consider the isotropic Coulomb interaction:
\begin{equation}
N(0) V_{\mathbf{k}-\mathbf{k}^{\prime}, n-m}=\lambda_{n,m} \Theta\left(k_c-|\mathbf{k}|\right) \Theta\left(k_c-|\mathbf{k}^{\prime}|\right),
\end{equation}
where $\lambda_{n,m}$ is dimensionless coupling and $\mathbf{k}_c = 2\mathbf{k}_F$ represents the momentum cutoff. The gap equation under isotropic approximation takes the form:
\begin{equation}
\phi_n=-2 \Tilde{T}_c \sum_{m=-\infty}^{\infty}  \frac{1}{Z_m \Tilde{\omega}_m} \arctan \frac{1}{Z_m  \Tilde{\omega}_m} \lambda_{n,m} \phi_m, 
\label{iso_gap}
\end{equation}
which is Eq.~(\ref{eq:iso_gap_eq}) in main text. In general, the function form of mass renormalization function $Z_n$ strongly depends on the form of coupling interaction $\lambda_{n,m}$, which needs to be determined numerically. Here we adopt the Lorentzian approximation with the Coulomb interaction described by Lorentzian form
\begin{equation}
\lambda_{n,m}=\mu\left(1- \frac{\Tilde{\Omega}_b^2}{\Tilde{\Omega}_b^2+(\Tilde{\omega}_n-\Tilde{\omega}_m)^2}\right),
\end{equation}
where we have defined thee dimension quantities $\Tilde{\Omega}_b=\Omega_b/E_F$. In this case, Eq.~(\ref{1st_self_corr}) can be evaluated analytically
\begin{equation}
\begin{aligned}
R_{\mathbf{k}, n}
&=-\mu\left(E_{\mathrm{F}}+i\Omega_b \arctan \frac{\Tilde{\omega}_n }{\Tilde{\omega}_n^2+\Tilde{\Omega}_b\left(1+\Tilde{\Omega}_b\right)}\right),
\end{aligned}
\end{equation}
which gives the solution of renormalization 
\begin{equation}
Z_n = 1+\mu\frac{\Tilde{\Omega}_b}{\Tilde{\omega}_n} \arctan \frac{\Tilde{\omega}_n }{\Tilde{\omega}_n^2+\Tilde{\Omega}_b\left(1+\Tilde{\Omega}_b\right)}.
\label{eq:ana_solution_Z}
\end{equation}

The procedures to determine critical temperature are the following: (i) Discrete the Brillouin zone and compute the band structures. We choose $100\times100$  mesh to ensure convergence. (ii) For a given temperature $T$, we generate Mastubara frequency grid by setting the energy cutoff $E_c=2000\ \mathrm{meV}$. (iii) The dynamic screened Coulomb interaction is therefore computed via Eq.~(\ref{average_coulomb}) and the coupling strength is given by Eq.~(\ref{eq:coupling_strength}). (iv) To compute critical temperature beyond Lorentzian approximation, we compute the mass renormalization $Z_m$ function numerically by integrating Eq.~(\ref{1st_self_corr}). (v) Gap equation (Eq.~(\ref{eq:iso_gap_eq})) is then solved by the power method to reduce the computational time. (vi) We repeat the above procedures and find the critical temperature $T_c$ when the largest eigenvalue is exactly equal to 1.

\section{Results for double-gate screening}\label{appendix:single_gate_screening}

Similar to single-gate screening, we consider the double-gate screening structure where the MATBG sample is placed between two external metal gates with the same separations $l$. The bare Coulomb interaction can be obtained in the limit of $\Pi_2,\kappa_t, \kappa_b\rightarrow\infty$:
\begin{equation}
V(\mathbf{q})=\frac{2\pi e^2}{\kappa_m q}\tanh{d_{g}q}.
\end{equation}

Figure~\ref{fig:tc_ext_gate_sdb}(a) shows the critical temperature as a function of the filling factor $\nu$ for various separation distances $l$. Similar to single-gate screening, the bimodal feature disappears and is replaced by a single peak near the Van Hove singularity as $l$ decreases. We found that double metal gates will provide a lower critical temperature due to the extra screening from other metal gates. In Fig.~\ref{fig:tc_ext_gate_sdb}(b), a non-monotonic transition is also observed for fixed filling factor cases. This non-monotonicity can also be understood by mapping to the Lorentzian model, as shown in Fig.~\ref{fig:tc_ext_gate_sdb}(c) and (d). Again, we noted that the critical temperature remains relatively constant across a wide range of separations but experiences a significant drop when $l\lesssim 3 \mathrm{nm}$.

\section{Bare Coulomb potential}\label{appendix:Bare_Coulomb}

Here we derive the general form of Coulomb interaction for the hybrid structure mentioned in the main text. The bare Coulomb interaction can be obtained by solving the displacement field $\bm{D}$ to satisfy
\begin{equation}
\nabla\cdot\bm{D}=0
\end{equation}
everywhere in space. Here, the displacement field is related to the electric field by $\bm{D}=\kappa \bm{E}$. The material in each slab divides the space into several parts
\begin{equation}
\begin{cases}
\kappa_{1} & ;\text{ for }z>d_1\\
\kappa_{2} & ;\text{ for }0<z<d_1\\
\kappa_{3} & ;\text{ for }-l<z<0\\
\kappa_{4} & ;\text{ for }-l-d_2<z<-l\\
\kappa_{5} & ;\text{ for }z<-l-d_2\\
\end{cases},
\end{equation}
where we have set the 2D layer at $z=0$ and the MATBG at $z=-l$. The corresponding Poisson equation can be obtained by invoking the relation between electric field and electric potential $\bm{E}=-\nabla \phi$
\begin{equation}
\begin{aligned}
\kappa_{i}\nabla^{2}\phi_{i}&=0;i=1,2,3,4,5.\\
\end{aligned}
\end{equation}

The solution of the Poisson equation is given by the form of
\begin{equation}
\begin{cases}
\phi_{1}\left(\bm{r},z\right)=A e^{i\bm{q}\cdot\bm{r}}e^{-qz} & ;z>d_1\\
\phi_{2}\left(\bm{r},z\right)=e^{i\bm{q}\cdot\bm{r}}(Be^{qz}+Ce^{-qz}) & ;0<z<d_1\\
\phi_{3}\left(\bm{r},z\right)=e^{i\bm{q}\cdot\bm{r}}(De^{qz}+Ee^{-qz}) & ;-l<z<0\\
\phi_{4}\left(\bm{r},z\right)=e^{i\bm{q}\cdot\bm{r}}(Fe^{qz}+Ge^{-qz}) & ;-l-d_2<z<-l\\
\phi_{5}\left(\bm{r},z\right)=H e^{i\bm{q}\cdot\bm{r}}e^{qz} & ;z<-l-d_2\\
\end{cases},
\end{equation}
where we have use shorthand $\bm{r}=\begin{pmatrix}x\\
y
\end{pmatrix}$, $\bm{q}=\begin{pmatrix}q_{x}\\
q_{y}
\end{pmatrix},$ and $q=|\bm{q}|$. This form has been chosen such that the potential doesn't diverge towards $z=\pm\infty$. The undetermined coefficients can be obtained by matching the electric potential at the boundary between each layer/slab~\cite{van2016plasmonics}. The solution of the Poisson equation can be written down as a linear equation
\begin{widetext}
\begin{equation}
\begin{pmatrix}
   e^{-qd_1} & -e^{qd_1} & -e^{-qd_1} & 0 & 0 & 0 & 0 & 0 \\
   0 & 1 & 1 & -1 & -1 & 0 & 0 & 0 \\
   0 & 0 & 0 & e^{-ql} & e^{ql} & -e^{-ql} & -e^{ql} & 0 \\
   0 & 0 & 0 & 0 & 0 & e^{-q(l+d_2)} & e^{q(l+d_2)} & -e^{-q(l+d_2)} \\
   -\kappa_1e^{-qd_1} & -\kappa_2e^{qd_1} & \kappa_2e^{-qd_1} & 0 & 0 & 0 & 0 & 0 \\
   0 & \kappa_2 & -\kappa_2 & -\kappa_3 & \kappa_3 & 0& 0 & 0 \\
   0 & 0 & 0 & \kappa_3 e^{-ql} & -\kappa_3 e^{ql} & -\kappa_4 e^{-ql} & \kappa_4 e^{ql} & 0 \\
   0 & 0 & 0 & 0 & 0 & \kappa_4 e^{-q(l+d_2)} & -\kappa_4 e^{q(l+d_2)} & -\kappa_5 e^{-q(l+d_2)} \\
 \end{pmatrix}
 \begin{pmatrix}
   A \\ B \\ C \\ D \\ E \\ F \\ G \\ H
 \end{pmatrix}
 =
 \begin{pmatrix}
   0 \\ 0 \\ 0 \\ 0 \\ 0 \\ 4\pi e/q \\ 0 \\ 0
 \end{pmatrix},
\end{equation}
\end{widetext}
which can be inverted to give us the undetermined coefficients. The intralayer bare Coulomb potential for the first layer is given by $V_{11}=-e\phi_{3}(0,0)=-e(D+E)$, and the interlayer bare Coulomb potential is given by $V_{12}=-e\phi_{3}\left(0,-l\right)=-e(De^{-ql}+Ee^{ql})$.

Due to the symmetry of system, the intralayer bare Coulomb potential $V_{22}$ can be obtained from $V_{11}$ by interchanging $\kappa_1\leftrightarrow \kappa_5, \kappa_2\leftrightarrow \kappa_4$ and $d_1\leftrightarrow d_2$. Similarly, the interlayer bare Coulomb potential $V_{12}$ can be obtained from $V_{21}$ using the same procedure. Finally, we obtain the bare Coulomb potential matrix elements

\begin{equation}
\begin{aligned}
V_{11}(q)&=\frac{4 \pi  e^2 }{c_0 q}e^{2 l q} f_1 \left(\kappa_3 \left(e^{2 l q}+1\right) f_2+\kappa_4 \left(e^{2 l q}-1\right) f_3\right)\\
V_{12}(q)&=V_{21}(q)=\frac{8 \pi  e^2 }{c_0 q} e^{3 l q} \kappa_3  f_1 f_2\\
V_{22}(q)&=V_{11}(\kappa_1\leftrightarrow \kappa_5, \kappa_2\leftrightarrow \kappa_4, d_1\leftrightarrow d_2),\\
\end{aligned}
\end{equation}
where
\begin{equation}
\begin{aligned}
c_0 &= (\kappa_1+\kappa_2) (\kappa_2-\kappa_3) (\kappa_3-\kappa_4) (\kappa_4+\kappa_5) e^{2 q (d_1+d_2+l)} \\
&+ (\kappa_1+\kappa_2) (\kappa_2+\kappa_3) (\kappa_3+\kappa_4) (\kappa_4+\kappa_5) e^{2 q (d_1+d_2+2 l)} \\
&+ (\kappa_1+\kappa_2) (\kappa_2+\kappa_3) (\kappa_3-\kappa_4) (\kappa_4-\kappa_5) e^{2 q (d_1+2 l)} \\
&+ (\kappa_1+\kappa_2) (\kappa_2-\kappa_3) (\kappa_3+\kappa_4) (\kappa_4-\kappa_5) e^{2 q (d_1+l)} \\
&+ (\kappa_1-\kappa_2) (\kappa_2+\kappa_3) (\kappa_3-\kappa_4) (\kappa_4+\kappa_5) e^{2 q (d_2+l)} \\
&+ (\kappa_1-\kappa_2) (\kappa_2-\kappa_3) (\kappa_3+\kappa_4) (\kappa_4+\kappa_5) e^{2 q (d_2+2 l)} \\
&+ (\kappa_1-\kappa_2) (\kappa_2-\kappa_3) (\kappa_3-\kappa_4) (\kappa_4-\kappa_5) e^{4 l q} \\
&+ (\kappa_1-\kappa_2) (\kappa_2+\kappa_3) (\kappa_3+\kappa_4) (\kappa_4-\kappa_5) e^{2 l q}, \\
\end{aligned}
\end{equation}
and
\begin{equation}
\begin{aligned}
f_1 &= e^{2 d_1 q} (\kappa_1+\kappa_2)-\kappa_1+\kappa_2 \\
f_2 &= e^{2 d_2 q} (\kappa_4+\kappa_5)+\kappa_4-\kappa_5 \\
f_3 &= e^{2 d_2 q} (\kappa_4+\kappa_5)-\kappa_4+\kappa_5. \\
\end{aligned}
\end{equation}

In the case of $\kappa_2=\kappa_3=\kappa_4=\kappa_m,\kappa_1\rightarrow\kappa_t,\kappa_5\rightarrow\kappa_m,d_1=d_2=d$, we obtain the following solution
\begin{widetext}
\begin{equation}
\begin{aligned}
& V_{11}(q)=\frac{4 \pi e^2}{\kappa_m q} \frac{\left[\kappa_m \cosh (q d)+\kappa_t \sinh (q d)\right]\left\{\kappa_m \cosh [q(l+d)]+\kappa_b \sinh [q(l+d)]\right\}}{\left(\kappa_t+\kappa_b\right) \kappa_m \cosh [q(l+2 d)]+\left(\kappa_t \kappa_b+\kappa_m^2\right) \sinh [q(l+2 d)]} \\
& V_{12}(q)=V_{21}(q)=\frac{4 \pi e^2}{\kappa_m q} \frac{\left[\kappa_m \cosh (q d)+\kappa_b \sinh (q d)\right]\left[\kappa_m \cosh (q d)+\kappa_t \sinh (q d)\right]}{\left(\kappa_t+\kappa_b\right) \kappa_m \cosh [q(l+2 d)]+\left(\kappa_t \kappa_b+\kappa_m^2\right) \sinh [q(l+2 d)]} \\
& V_{22}(q)=\frac{4 \pi e^2}{\kappa_m q} \frac{\left[\kappa_m \cosh (q d)+\kappa_b \sinh (q d)\right]\left\{\kappa_m \cosh [q(l+d)]+\kappa_t \sinh [q(l+d)]\right\}}{\left(\kappa_t+\kappa_b\right) \kappa_m \cosh [q(l+2 d)]+\left(\kappa_t \kappa_b+\kappa_m^2\right) \sinh [q(l+2 d)]}
\end{aligned}.
\end{equation}
\end{widetext}

\section{Effective Coulomb interaction for MATBG-2D hybrid system}\label{appendix:2lRPA}

The Hamiltonian which describes MATBG-2D system can be written as
\begin{equation}
    \mathcal{H} = \mathcal{H}_{1}+\mathcal{H}_{2}+\mathcal{H}_{ee},
\end{equation}
where $H_1$ ($H_2$) indicates the non-interacting Hamiltonian of MATBG (2D material). The coupling Hamiltonian reads
\begin{equation}
\mathcal{H}_{\mathrm{ee}}=\frac{1}{2} \sum_{\boldsymbol{q}, \ell \neq \ell^{\prime}} V_{\ell \ell^{\prime}}(q) \hat{n}_{\boldsymbol{q}, \ell} \hat{n}_{-\boldsymbol{q}, \ell^{\prime}},
\end{equation}
where $\hat{n}_{\boldsymbol{q}, \ell}$ is the density operator of the $\ell$-th layer
\begin{equation}
\hat{n}_{\boldsymbol{q}, \ell}=\sum_{\boldsymbol{k}, \alpha} \hat{\psi}_{\boldsymbol{k}-\boldsymbol{q}, \alpha, \ell}^{\dagger} \hat{\psi}_{\boldsymbol{k}, \alpha, \ell},
\end{equation}
and $V_{\ell \ell^{\prime}}(q)$ ($\left.\ell \neq \ell^{\prime}\right)$ is the 2D Fourier transform of the inter-layer Coulomb interaction.

For convenience, we introduce the matrix
\begin{equation}
\hat{W}=\left(\begin{array}{ll}
W_{11} & W_{12} \\
W_{21} & W_{22}
\end{array}\right),
\hat{V}=\left(\begin{array}{ll}
V_{11} & V_{12} \\
V_{21} & V_{22}
\end{array}\right),
\end{equation}
to represent the screened and bare Coulomb interaction respectively. The details about bare Coulomb interaction can be found in Appendix~\ref{appendix:Bare_Coulomb}. From the random phase approximation, the screened Coulomb potential matrix $\hat{W}$ is given by
\begin{equation}
\begin{aligned}
\hat{W} & = \hat{V}+\hat{V} \hat{\Pi} \hat{V}+\hat{V} \hat{\Pi} \hat{V} \hat{\Pi} \hat{V}+\ldots \\
& = \hat{V}+\hat{V} \hat{\Pi} \hat{W},
\end{aligned}
\label{rpa}
\end{equation}
where $\hat{\Pi}$ is the polarizability matrix. Eq.~(\ref{rpa}) can be rearranged into
\begin{equation}
\begin{aligned}
\hat{W} 
& = \left( \hat{1} -\hat{V}\hat{\Pi} \right)^{-1}\hat{V} \\
& = \hat{\mathbf{\epsilon}}^{-1} \hat{V},
\end{aligned}
\label{rpa1}
\end{equation}
where $\hat{\mathbf{\epsilon}}$ is the dielectric matrix which can be evaluated to be $\hat{\epsilon}^{-1} =(\hat{1}-\hat{V} \hat{\Pi})^{-1}$. For the system with few nanometers' separations, the interlayer tunneling is negligible compare with intralayer coupling, we assume the off-diagonal term of $\hat{\Pi}$ is zero, i.e. $\Pi_{i j}=\Pi_i \delta_{i j}$. Under this assumption, the dielectric matrix is written as 
\begin{equation}
\begin{aligned}
\hat{\epsilon}^{-1} & =\left(\begin{array}{cc}
1-V_{11} \Pi_1 & -V_{12} \Pi_2 \\
-V_{21} \Pi_1 & 1-V_{22} \Pi_2
\end{array}\right)^{-1} \\
& =\frac{1}{\epsilon_1 \epsilon_2-V_{12} V_{21} \Pi_1 \Pi_2}\left(\begin{array}{cc}
\epsilon_2 & V_{12} \Pi_2 \\
V_{21} \Pi_1 & \epsilon_1
\end{array}\right),
\end{aligned}
\label{dielectric}
\end{equation}
where we have defined dielectric function $\epsilon_i=1-V_{i i} \Pi_i$. Combine Eq.~(\ref{rpa1}) and Eq.~(\ref{dielectric}), we obtain
\begin{equation}
\hat{W}= \frac{1}{\epsilon_1 \epsilon_2-V_{12} V_{21} \Pi_1 \Pi_2}
\left(\begin{array}{cc} 
\epsilon_2 & V_{12} \Pi_2 \\
V_{21} \Pi_1 & \epsilon_1
\end{array}\right) 
\left(\begin{array}{ll}
V_{11} & V_{12} \\
V_{21} & V_{22}
\end{array}\right).
\end{equation}

We are particularly interested in the Coulomb interaction in the active layer, which corresponds to the $W_{11}$ element
\begin{equation}
V(\mathbf{q}, i \omega)=W_{11}=\frac{\epsilon_{2}V_{11}+V_{12}V_{21}\Pi_{2}}{\epsilon_1 \epsilon_2-V_{12} V_{21} \Pi_1 \Pi_2}.
\label{eq:pairing_element1}
\end{equation}

Eq.~(\ref{eq:pairing_element1}) can be rewritten as more compact form
\begin{equation}
V(\mathbf{q}, i \omega)=\frac{V_{\mathrm{eff}}}{1-\Pi_1 V_{\mathrm{eff}}}, \\
\end{equation}
where we have defined the effective bare Coulomb interaction as
\begin{equation}
V_{\mathrm{eff}}(\mathbf{q}, i \omega)=V_{11}\left[1-\frac{V_{12}V_{21}}{V_{11}V_{22}}\left(1-\frac{1}{\epsilon_2}\right)\right],
\label{eq:eff_bare_Coulomb1}
\end{equation}
which is the Eq.~(\ref{eq:eff_bare_Coulomb}) in the main text.

In the case of bare Coulomb interaction
\begin{equation}
V_{11}=V_{22}=\frac{2\pi e^2}{q}; V_{12}=V_{21}=\frac{2\pi e^2}{q} e^{-q l},
\end{equation}
where $l$ is the distance between the active and passive layers. Eq.~(\ref{eq:eff_bare_Coulomb1}) then become
\begin{equation}
V_{\mathrm{eff}} =\frac{2\pi e^2}{q}\left[1-e^{-2ql}\left(1-\frac{1}{\epsilon_2}\right)\right],
\end{equation}
which is exactly the Coulomb interaction derived in~\cite{Xiaoxue_Science2021_BLG}.

\bibliographystyle{apsrev4-2}
\bibliography{refs}

\end{document}